\documentclass[notitlepage,11pt]{article}%
\usepackage{amsfonts}
\usepackage{amsmath}
\usepackage{amssymb}
\usepackage{graphicx}%
\newtheorem{theorem}{Theorem}
\newtheorem{acknowledgement}[theorem]{Acknowledgement}

\newtheorem{corollary}[theorem]{Corollary}

\newtheorem{lemma}[theorem]{Lemma}

\newtheorem{proposition}[theorem]{Proposition}
\newtheorem{remark}[theorem]{Remark}

\newenvironment{proof}[1][Proof]{\noindent\textbf{#1.} }{\ \rule{0.5em}{0.5em}}
\begin{document}

\title{Extension of the Conley--Zehnder Index and Calculation of the Maslov-Type
Index Intervening in Gutzwiller's Trace Formula}
\author{Maurice de Gosson\\{\small Universit\"{a}t Potsdam, Inst. f. Mathematik}\\{\small Am Neuen Palais 10, D-14415 Potsdam}\\{\small Current address:}\\{\small Universidade de S\~{a}o Paulo, Dep. de Matem\~{a}tica}\\{\small CEP 05508-900 S\~{a}o Paulo}\\{\small E-mail: maurice.degosson@gmail}\texttt{.com}
\and Serge de Gosson\\{\small V\"{a}xj\"{o} University, MSI, }\\{\small V\"{a}xj\"{o}, SE-351 95}\\{\small E-mail: sergedegosson@gmail.com}}
\maketitle

\begin{abstract}
The aim of this paper is to define and study a non-trivial extension of the
Conley--Zehnder index. We make use for this purpose of the global
Arnol'd--Leray--Maslov index constructed by the authors in previous work. This
extension allows us to obtain explicit formulae for the calculation of the
Gutzwiller--Maslov index of a Hamiltonian periodic orbit. We show that this
index is related to the usual Maslov index via Morse's index of concavity of a
periodic Hamiltonian orbit. In addition we prove a formula allowing to
calculate the index of a repeated orbit.

\end{abstract}

\bigskip


\textit{AMS\ Classification} (2000): 37J05, 53D12, 53Z05, 81S30

\section{Introduction}

This paper is devoted to a redefinition and extension of the Conley--Zehnder
index \cite{CZ,HWZ} from the theory of periodic Hamiltonian orbits. This will
allow us to prove two new (and useful) formulae. The first of these formulae
expresses the index of the product of two symplectic paths in terms of the
\emph{symplectic Cayley transform} of a symplectic matrix; the second shows
that the difference between the Conley--Zehnder index and the usual Maslov
index is (up to the sign) \textit{Morse's index of concavity}.

Besides the intrinsic interest of our constructions (they allow practical
calculations of the Conley--Zehnder index even in degenerate cases) they
probably settle once for all some well-known problems from Gutzwiller's
\cite{Gutzwiller} quantization scheme of classically chaotic systems. Let us
briefly recall the idea. Let $\widehat{H}$ be a Hamiltonian operator with
discrete spectrum $E_{1},E_{2},...$. One wants to find an asymptotic
expression, for $\hbar\rightarrow0$, of the \textquotedblleft level
density\textquotedblright\
\[
d(E)=\sum_{k=1}^{\infty}\delta(E-E_{j})=-\frac{1}{\pi}\lim_{\varepsilon
\rightarrow0+}\operatorname*{Tr}G(x,x^{\prime},E+i\varepsilon)
\]
($G$ the Green function determined by $\widehat{H}$). Writing $d(E)=\bar
{d}(E)+\widetilde{d}(E)$ where $\bar{d}(E)$ is the Thomas--Fermi or smoothed
density of states,\ and $\widetilde{d}(E)$ the \textquotedblleft oscillating
term\textquotedblright, Gutzwiller's formula says that in the limit
$\hbar\rightarrow0$ we have
\[
\widetilde{d}(E)=\widetilde{d}_{\text{Gutz}}(E)+O(\hbar)
\]
where
\begin{equation}
\widetilde{d}_{\text{Gutz}}(E)=\frac{1}{\pi\hbar}\operatorname{Re}\sum
_{\gamma}\frac{T_{\gamma}i^{\mu_{\gamma}}}{\sqrt{|\det(\tilde{S}_{\gamma}%
-I)|}}\exp\left(  \frac{i}{\hbar}\oint\nolimits_{\gamma}pdx\right)  \text{.}
\label{Gutz}%
\end{equation}
The sum in the right-hand side of the formula above is taken over all
\textit{periodic orbits} $\gamma$ with period $T_{\gamma}$ of the classical
Hamiltonian $H$, \emph{including their repetitions}; the exponent $\mu
_{\gamma}$ is an integer and and $\tilde{S}_{\gamma}$ is the stability matrix
of $\gamma$. Gutzwiller's theory is far from being fully understood (there are
problems due to possible divergences of the series, insufficient error
estimates, etc.). We will not discuss these delicate problems here; what we
will do is instead to focus on the integers $\mu_{\gamma}$, to which much
literature has been devoted (see for instance \cite{BB,CRL,Robbins} and the
references therein). As is well-known, $\mu_{\gamma}$ is not the usual Maslov
index familiar from EBK quantization of Lagrangian manifolds, but rather (up
to the sign) the Conley--Zehnder index of the orbit $\gamma$:%
\[
``\text{Gutzwiller--Maslov index\textquotedblright}=-(\text{Conley--Zehnder
index})
\]
(see \textit{e.g.} Muratore-Ginnaneschi in \cite{MG}; in \cite{mdglettmath}
one of us has given a different proof using an ingenious and promising
derivation of Gutzwiller's formula by Mehlig and Wilkinson \cite{WM} based on
the Weyl representation of metaplectic operators). We notice that Sugita
\cite{Sugita} has proposed a scheme for the calculation of what is essentially
the Conley--Zehnder index using a Feynman path-integral formalism; his
constructions are however mathematically illegitimate and seem difficult to
justify without using deep results from functional analysis (but the effort
might perhaps be worthwhile).\bigskip

This paper is structured as follows:

\begin{enumerate}
\item Redefine the Conley--Zehnder index in terms of globally defined indices
(the Wall--Kashiwara index and the Arnol'd--Leray--Maslov index); we will thus
obtain a non-trivial extension $\nu$ of that index which is explicitly
computable for all paths, even in the case $\det(S-I)=0$; this is useful in
problems where degeneracies arise (for instance the isotropic harmonic
oscillator, see \cite{BP} and the example of last section in the present paper);

\item We will prove a simple formula for the Conley--Zehnder index of the
product of two symplectic paths. We will see that in particular the index of
an orbit which is repeated $r$ times is%
\[
\nu_{\gamma_{r}}=r\nu_{\gamma}+\tfrac{1}{2}(r-1)\operatorname*{sign}M_{S}%
\]
where
\[
M_{S}=\frac{1}{2}J(S+I)(S-I)^{-1}%
\]
is the symplectic Cayley transform of the monodromy matrix $S$;

\item We will finally prove that the Conley--Zehnder index $\nu_{\gamma}$ of a
non-degenerate periodic orbit $\gamma$ is related to its Maslov index
$m_{\gamma}$ by the simple formula%
\[
\nu_{\gamma}=m_{\gamma}-\operatorname*{Inert}W_{\gamma}%
\]
where $\operatorname*{Inert}W_{\gamma}$ is Morse's \textquotedblleft index of
concavity\textquotedblright\ \cite{Morse,MG}, defined in terms of the
generating function of the monodromy matrix.
\end{enumerate}

We close this article by performing explicit calculations in the case of the
two-dimensional anisotropic harmonic oscillator; this allows us to recover a
formula obtained by non-rigorous methods in the literature.

\subsection*{Notations}

We will denote by $\sigma$ the standard symplectic form on $\mathbb{R}%
^{2n}=\mathbb{R}_{x}^{n}\times\mathbb{R}_{p}^{n}$:%
\[
\sigma(z,z^{\prime})=\left\langle p,x^{\prime}\right\rangle -\left\langle
p^{\prime},x\right\rangle \text{ \ if \ }z=(x,p)\text{, }z^{\prime}%
=(x^{\prime}p^{\prime})
\]
that is, in matrix form%
\[
\sigma(z,z^{\prime})=\left\langle Jz,z^{\prime}\right\rangle \text{ \ \ ,
\ \ }J=%
\begin{bmatrix}
0 & I\\
-I & 0
\end{bmatrix}
\text{.}%
\]
The real symplectic group $\operatorname*{Sp}(n)$ consists of all linear
automorphisms $S$ of $\mathbb{R}^{2n}$ such that $\sigma(Sz,Sz^{\prime
})=\sigma(z,z^{\prime})$ for all $z,z^{\prime}$. Equivalently:%
\[
S\in\operatorname*{Sp}(n)\Longleftrightarrow S^{T}JS=SJS^{T}=J\text{.}%
\]
$\operatorname*{Sp}(n)$ is a connected Lie group and $\pi_{1}%
[\operatorname*{Sp}(n)]\equiv(\mathbb{Z},+)$. We denote by $\operatorname{Lag}%
(n)$ the Lagrangian Grassmannian of $(\mathbb{R}^{2n},\sigma)$, that is:
$\ell\in\operatorname{Lag}(n)$ if and only $\ell$ is a $n$-plane in
$\mathbb{R}^{2n}$ on which $\sigma$ vanishes identically. We will write
$\ell_{X}=\mathbb{R}_{x}^{n}\times0$ and $\ell_{P}=0\times\mathbb{R}_{p}^{n}$
(the \textquotedblleft horizontal\textquotedblright\ and \textquotedblleft
vertical\textquotedblright\ polarizations).

\section{Prerequisites}

In this section we review previous results \cite{JMPA,Wiley,MSDG} on
Lagrangian and symplectic Maslov indices generalizing those of Leray
\cite{Leray}. An excellent comparative study of the indices used here with
other indices appearing in the literature can be found in Cappell \textit{et
al}. \cite{CLM}.

In what follows $(E,\omega)$ is a finite-dimensional symplectic space, $\dim
E=2n$, and $\operatorname{Sp}(E,\omega)$, $\operatorname*{Lag}(E,\omega)$ the
associated symplectic group and Lagrangian Grassmannian.

\subsection{The Wall--Kashiwara index}

Let $(\ell,\ell^{\prime},\ell^{\prime\prime})$ be a triple of elements of
$\operatorname*{Lag}(E,\omega)$; by definition \cite{CLM,LV,Wall} the
Wall--Kashiwara index (or: signature) $\tau(\ell,\ell^{\prime},\ell
^{\prime\prime})$ is the signature of the quadratic form
\[
Q(z,z^{\prime},z^{\prime\prime})=\sigma(z,z^{\prime})+\sigma(z^{\prime
},z^{\prime\prime})+\sigma(z^{\prime\prime},z^{\prime})
\]
on $\ell\oplus\ell^{\prime}\oplus\ell^{\prime\prime}$. The index $\tau$ is
antisymmetric:
\[
\tau(\ell,\ell^{\prime},\ell^{\prime\prime})=-\tau(\ell^{\prime},\ell
,\ell^{\prime\prime})=-\tau(\ell,\ell^{\prime\prime},\ell^{\prime})=-\tau
(\ell^{\prime\prime},\ell^{\prime},\ell);
\]
it is a symplectic invariant:%
\[
\tau(S\ell,S\ell^{\prime},S\ell^{\prime\prime})=\tau(\ell,\ell^{\prime}%
,\ell^{\prime\prime})\text{ \ for }S\in\operatorname{Sp}(n)
\]
and it has the following essential cocycle property:%
\begin{equation}
\tau(\ell,\ell^{\prime},\ell^{\prime\prime})-\tau(\ell^{\prime},\ell
^{\prime\prime},\ell^{\prime\prime\prime})+\tau(\ell^{\prime},\ell
^{\prime\prime},\ell^{\prime\prime\prime})-\tau(\ell^{\prime},\ell
^{\prime\prime},\ell^{\prime\prime\prime})=0\text{.} \label{coka}%
\end{equation}
Moreover its values modulo $2$ are given by the formula:%

\begin{equation}
\tau(\ell,\ell^{\prime},\ell^{\prime\prime})\equiv n+\dim\ell\cap\ell^{\prime
}+\dim\ell^{\prime}\cap\ell^{\prime\prime}+\dim\ell^{\prime\prime}\cap
\ell\text{ \ }\operatorname{mod}2\text{.} \label{kmod2}%
\end{equation}
Let $(E,\omega)=(E^{\prime}\oplus E^{\prime\prime},\omega^{\prime}\oplus
\omega^{\prime\prime})$; identifying $\operatorname*{Lag}(E^{\prime}%
,\omega^{\prime})\oplus\operatorname*{Lag}(E^{\prime\prime},\omega
^{\prime\prime})$ with a subset of $\operatorname*{Lag}(E,\omega)$ we have the
following additivity formula:%
\[
\tau(\ell_{1}\oplus\ell_{2},\ell_{1}^{\prime}\oplus\ell_{2}^{\prime},\ell
_{1}^{\prime\prime}\oplus\ell_{2}^{\prime\prime})=\tau_{1}(\ell_{1},\ell
_{1}^{\prime},\ell_{1}^{\prime\prime})+\tau_{2}(\ell_{2},\ell_{2}^{\prime
},\ell_{2}^{\prime\prime})
\]
where $\tau_{1}$ and $\tau_{2}$ are the signatures on $\operatorname*{Lag}%
(E^{\prime},\omega^{\prime})$ and $\operatorname*{Lag}(E^{\prime\prime}%
,\omega^{\prime\prime})$.

The following Lemma will be helpful in our study of the Conley--Zehnder index:

\begin{lemma}
\label{lemsimple}(i) If $\ell\cap\ell^{\prime\prime}=0$ then $\tau(\ell
,\ell^{\prime},\ell^{\prime\prime})$ is the signature of the quadratic form
\[
Q^{\prime}(z^{\prime})=\omega(\Pr\nolimits_{\ell\ell^{\prime\prime}}z^{\prime
},z^{\prime})=\omega(z^{\prime},\Pr\nolimits_{\ell^{\prime\prime}\ell
}z^{\prime})
\]
on $\ell^{\prime}$, where $\Pr\nolimits_{\ell\ell^{\prime\prime}}$ is the
projection onto $\ell$ along $\ell^{\prime\prime}$ and $\Pr\nolimits_{\ell
^{\prime\prime}\ell}=I-\Pr\nolimits_{\ell\ell^{\prime\prime}}$ is the
projection on $\ell^{\prime\prime}$ along $\ell$. (ii) Let $(\ell,\ell
^{\prime},\ell^{\prime\prime})$ be a triple of Lagrangian planes such that an
$\ell=\ell\cap\ell^{\prime}+\ell\cap\ell^{\prime\prime}$. Then $\tau(\ell
,\ell^{\prime},\ell^{\prime\prime})=0$.
\end{lemma}

\noindent(See \textit{e.g.} \cite{LV} for a proof).

The index of inertia of the triple $(\ell,\ell^{\prime},\ell^{\prime\prime})$
is defined by%
\begin{equation}
\operatorname*{Inert}(\ell,\ell^{\prime},\ell^{\prime\prime})=\frac{1}{2}%
(\tau(\ell,\ell^{\prime},\ell^{\prime\prime})+n+\dim\ell\cap\ell^{\prime}%
-\dim\ell^{\prime}\cap\ell^{\prime\prime}+\dim\ell^{\prime\prime}\cap
\ell)\text{; } \label{defi}%
\end{equation}
in view of (\ref{kmod2}) it is an integer. When the Lagrangian planes $\ell$,
$\ell^{\prime}$, $\ell^{\prime\prime}$ are pairwise transverse it follows from
the first part of Lemma \ref{lemsimple} that $\operatorname*{Inert}(\ell
,\ell^{\prime},\ell^{\prime\prime})$ coincides with the index of inertia
defined by Leray \cite{Leray}: see \cite{JMPA,Wiley}.

\subsection{The ALM index}

Recall \cite{JMPA,Wiley} (also \cite{MSDG} for a review and \cite{ICP} for
calculations in the case $n=1$) that the ALM (=Arnol'd--Leray--Maslov) index
on the universal covering $\operatorname{Lag}_{\infty}(E,\omega)$ of
$\operatorname{Lag}(E,\omega)$ is the unique mapping%
\[
\mu:(\operatorname{Lag}_{\infty}(E,\omega))^{2}\longrightarrow\mathbb{Z}%
\]
having the two following properties:

\begin{itemize}
\item $\mu$ is locally constant on each set $\{(\ell_{\infty},\ell_{\infty
}^{\prime}):\dim\ell\cap\ell^{\prime}=k\}$ ($0\leq k\leq n$);

\item For all $\ell_{\infty}$, $\ell_{\infty}^{\prime}$, $\ell_{\infty
}^{\prime\prime}$ in $\operatorname{Lag}_{\infty}(E,\omega)$ with projections
$\ell$, $\ell^{\prime}$, $\ell^{\prime\prime}$ we have
\begin{equation}
\mu(\ell_{\infty},\ell_{\infty}^{\prime})-\mu(\ell_{\infty},\ell_{\infty
}^{\prime\prime})+\mu(\ell_{\infty}^{\prime},\ell_{\infty}^{\prime\prime
})=\tau(\ell,\ell^{\prime},\ell^{\prime\prime}) \label{un}%
\end{equation}
where $\tau$ is the Wall--Kashiwara index on $\operatorname{Lag}(E,\omega)$.
\end{itemize}

The ALM\ index has in addition the following properties:%
\begin{equation}
\mu(\ell_{\infty},\ell_{\infty}^{\prime})\equiv n+\dim\ell\cap\ell^{\prime
}\text{ \ \ }\operatorname{mod}2 \label{mod2}%
\end{equation}
($n=\frac{1}{2}\dim E$) and%

\begin{equation}
\mu(\beta^{r}\ell_{\infty},\beta^{r^{\prime}}\ell_{\infty}^{\prime})=\mu
(\ell_{\infty},\ell_{\infty}^{\prime})+2(r-r^{\prime}) \label{trois}%
\end{equation}
for all integers $r$ and $r^{\prime}$; here $\beta$ denotes the generator of
$\pi_{1}[\operatorname{Lag}(E,\omega)]\equiv(\mathbb{Z},+)$ whose image in
$\mathbb{Z}$ is $+1$. From the dimensional additivity property of the
signature $\tau$ immediately follows that if $\ell_{1,\infty}\oplus
\ell_{2,\infty}$ and $\ell_{1,\infty}^{\prime}\oplus\ell_{2,\infty}^{\prime}$
are in
\[
\operatorname*{Lag}\nolimits_{\infty}(E^{\prime},\omega^{\prime}%
)\oplus\operatorname*{Lag}\nolimits_{\infty}(E^{\prime\prime},\omega
^{\prime\prime})\subset\operatorname*{Lag}\nolimits_{\infty}(E,\omega)
\]
then%
\begin{equation}
\mu(\ell_{1,\infty}\oplus\ell_{2,\infty},\ell_{1,\infty}^{\prime}\oplus
\ell_{2,\infty}^{\prime})=\mu^{\prime}(\ell_{1,\infty},\ell_{1,\infty}%
^{\prime})+\mu^{\prime\prime}(\ell_{2,\infty},\ell_{2,\infty}^{\prime})
\label{almad}%
\end{equation}
where $\mu^{\prime}$ and $\mu^{\prime\prime}$ are the ALM indices on
$\operatorname*{Lag}\nolimits_{\infty}(E^{\prime},\omega^{\prime})$ and
$\operatorname*{Lag}\nolimits_{\infty}(E^{\prime\prime},\omega^{\prime\prime
})$, respectively.

When $(E,\omega)$ is the standard symplectic space $(\mathbb{R}^{2n},\sigma)$
the \textquotedblleft Souriau mapping\textquotedblright\ \cite{Souriau}
identifies $\operatorname*{Lag}(E,\omega)=\operatorname{Lag}(n)$ with the set%
\[
\operatorname*{W}(n,\mathbb{C})=\{w\in\operatorname*{U}(n,\mathbb{C}%
):w=w^{T}\}
\]
of symmetric unitary matrices. This is done by associating to to $\ell
=u\ell_{P}$ ($u\in\operatorname*{U}(n,\mathbb{C})$) the matrix $w=uu^{T}$; the
Maslov bundle $\operatorname{Lag}_{\infty}(n)$ is then identified with%
\[
W_{\infty}(n,\mathbb{C})=\{(w,\theta):w\in\operatorname*{W}(n,\mathbb{C}%
)\text{, }\det w=e^{i\theta}\};
\]
the projection $\pi^{\operatorname{Lag}}:\ell_{\infty}\longmapsto\ell$
becoming $(w,\theta)\longmapsto w$. The ALM\ index is then calculated as
follows:\bigskip

\begin{itemize}
\item If $\ell\cap\ell^{\prime}=0$ then
\begin{equation}
\mu(\ell_{\infty},\ell_{\infty}^{\prime})=\frac{1}{\pi}\left[  \theta
-\theta^{\prime}+i\operatorname*{Tr}\operatorname{Log}(-w(w^{\prime}%
)^{-1}\right]  \label{souriau}%
\end{equation}
(the transversality condition $\ell\cap\ell^{\prime}$ is equivalent to
$-w(w^{\prime})^{-1}$ having no negative eigenvalue);\bigskip

\item If $\ell\cap\ell^{\prime}\neq0$ one chooses any $\ell^{\prime\prime}$
such that $\ell\cap\ell^{\prime\prime}=\ell^{\prime}\cap\ell^{\prime\prime}=0$
and one then calculates $\mu(\ell_{\infty},\ell_{\infty}^{\prime})$ using the
formula (\ref{un}), the values of $\mu(\ell_{\infty},\ell_{\infty}%
^{\prime\prime})$ and $\mu(\ell_{\infty}^{\prime},\ell_{\infty}^{\prime\prime
})$ being given by (\ref{souriau}). (The cocycle property (\ref{coka}) of
$\tau$ guarantees that the result does not depend on the choice of
$\ell^{\prime\prime}$, see \cite{JMPA,Wiley}).
\end{itemize}

\subsection{The relative Maslov indices on $\operatorname*{Sp}(E,\omega
)$\label{relma}}

We begin by recalling the definition of the Maslov index for loops in
$\operatorname*{Sp}(n)$. Let $\gamma$ be a continuous mapping
$[0,1]\longrightarrow\operatorname*{Sp}(n)$ such that $\gamma(0)=\gamma(1)$,
and set $\gamma(t)=S_{t}$. Then $U_{t}=(S_{t}S_{t})^{-1/2}S_{t}$ is the
orthogonal part in the polar decomposition of $S_{t}$:
\[
U_{t}\in\operatorname*{Sp}(n)\cap\operatorname*{O}(2n,\mathbb{R})\text{.}%
\]
Let us denote by $u_{t}$ the image $\iota(U_{t})$ of $U_{t}$ in
$\operatorname*{U}(n,\mathbb{C})$:%
\[
\iota(U_{t})=A+iB\text{ \ if \ }U=%
\begin{bmatrix}
A & -B\\
B & A
\end{bmatrix}
\]
and define $\rho(S_{t})=\det u_{t}$. The Maslov index of $\gamma$ is by
definition the degree of the loop $t\longmapsto\rho(S_{t})$ in $S^{1}$:%
\[
m(\gamma)=\deg[t\longmapsto\det(\iota(U_{t}))]\text{ , }0\leq t\leq1\text{.}%
\]
Let $\alpha$ be the generator of $\pi_{1}[\operatorname{Sp}(E,\omega
)]\equiv(\mathbb{Z},+)$ whose image in $\mathbb{Z}$ is $+1$; if $\gamma$ is
homotopic to $\alpha^{r}$ then
\begin{equation}
m(\gamma)=m(\alpha^{r})=2r\text{.} \label{masalfa}%
\end{equation}

The definition of the Maslov index can be extended to arbitrary paths in
$\operatorname{Sp}(E,\omega)$ using the properties of the ALM\ index. This is
done as follows: let $\ell=\pi^{\operatorname*{Lag}}(\ell_{\infty}%
)\in\operatorname*{Lag}(E,\omega)$; we define the Maslov index of $S_{\infty
}\in\operatorname{Sp}_{\infty}(E,\omega)$ relative to $\ell$ by%
\begin{equation}
\mu_{\ell}(S_{\infty})=\mu(S_{\infty}\ell_{\infty},\ell_{\infty})\text{;}
\label{mudef}%
\end{equation}
one shows (see \cite{JMPA,Wiley}) that the right-hand side only depends on the
projection $\ell$ of $\ell_{\infty}$, justifying the notation.\medskip

Here are three fundamental properties of the relative Maslov index; we will
need all of them to study the Conley--Zehnder index:

\begin{itemize}
\item \textit{Product}: For all $S_{\infty}$, $S_{\infty}^{\prime}$ in
$\operatorname{Sp}_{\infty}(E,\omega)$ we have%
\begin{equation}
\mu_{\ell}(S_{\infty}S_{\infty}^{\prime})=\mu_{\ell}(S_{\infty})+\mu_{\ell
}(S_{\infty}^{\prime})+\tau(\ell,S\ell,SS^{\prime}\ell); \label{uno}%
\end{equation}

\item \textit{Action of} $\pi_{1}[\operatorname{Sp}(n)]$: We have%
\begin{equation}
\mu_{\ell}(\alpha^{r}S_{\infty})=\mu_{\ell}(S_{\infty})+4r \label{sex}%
\end{equation}
for all $r\in\mathbb{Z}$;

\item \textit{Topological property}: The mapping $(S_{\infty},\ell
)\longmapsto\mu_{\ell}(S_{\infty})$ is locally constant on each of the sets
\begin{equation}
\{(S_{\infty},\ell):\dim S\ell\cap\ell=k\}\subset\operatorname{Sp}_{\infty
}(E,\omega)\times\operatorname*{Lag}(E,\omega) \label{splag}%
\end{equation}
($0\leq k\leq n$).
\end{itemize}

The two first properties readily follow from, respectively, (\ref{un}) and
(\ref{trois}). The third follows from the fact that the ALM\ index is locally
constant on the sets $\{(\ell_{\infty},\ell_{\infty}^{\prime}):\dim\ell
\cap\ell^{\prime}=k\}$. Note that (\ref{sex}) implies that%
\[
\mu_{\ell}(\alpha^{r})=4r
\]
hence the restriction of any of the $\mu_{\ell}$ to loops $\gamma$ in
$\operatorname{Sp}(E,\omega)$ is \emph{twice} the Maslov index $m(\gamma)$
defined above; it is therefore sometimes advantageous to use the variant of
$\mu_{\ell}$ defined by%
\begin{equation}
m_{\ell}(S_{\infty})=\frac{1}{2}(\mu_{\ell}(S_{\infty})+n+\dim(S\ell\cap\ell))
\label{remas}%
\end{equation}
where $n=\frac{1}{2}\dim E$. We will call $m_{\ell}(S_{\infty})$ the
\emph{reduced} (relative) Maslov index. In view of property (\ref{mod2}) it is
an integer; the properties of $m_{\ell}$ are obtained, \textit{mutatis
mutandis}, from those of $\mu_{\ell}$; for instance property (\ref{uno})
becomes%
\[
m_{\ell}(S_{\infty}S_{\infty}^{\prime})=m_{\ell}(S_{\infty})+m_{\ell
}(S_{\infty}^{\prime})+\operatorname*{Inert}(\ell,S\ell,SS^{\prime}\ell)
\]
where $\operatorname*{Inert}$ is the index of inertia defined by (\ref{defi}).

It follows from the cocycle property of the signature $\tau$ that the Maslov
indices corresponding to two choices $\ell$ and $\ell^{\prime}$ are related by
the formula%
\begin{equation}
\mu_{\ell}(S_{\infty})-\mu_{\ell^{\prime}}(S_{\infty})=\tau(S\ell,\ell
,\ell^{\prime})-\tau(S\ell,S\ell^{\prime},\ell^{\prime}); \label{sl}%
\end{equation}
similarly%
\begin{equation}
m_{\ell}(S_{\infty})-m_{\ell^{\prime}}(S_{\infty})=\operatorname*{Inert}%
(S\ell,\ell,\ell^{\prime})-\operatorname*{Inert}(S\ell,S\ell^{\prime}%
,\ell^{\prime})\text{.} \label{slm}%
\end{equation}

Assume that $(E,\omega)=(E^{\prime}\oplus E^{\prime\prime},\omega^{\prime
}\oplus\omega^{\prime\prime})$ and $\ell^{\prime}\in\operatorname*{Lag}%
(E^{\prime},\omega^{\prime})$, $\ell^{\prime\prime}\in\operatorname*{Lag}%
(E^{\prime\prime},\omega^{\prime\prime})$; the additivity property
(\ref{almad}) of the ALM\ index implies that if $S_{\infty}^{\prime}%
\in\operatorname{Sp}_{\infty}(E^{\prime},\omega^{\prime})$, $S_{\infty
}^{\prime\prime}\in\operatorname{Sp}_{\infty}(E^{\prime\prime},\omega
^{\prime\prime})$ then
\begin{equation}
\mu_{\ell^{\prime}\oplus\ell^{\prime\prime}}(S_{\infty}^{\prime}\oplus
S_{\infty}^{\prime\prime})=\mu_{\ell^{\prime}}(S_{\infty}^{\prime})+\mu
_{\ell_{2}}(S_{\infty}^{\prime\prime}) \label{masad}%
\end{equation}
where $\operatorname{Sp}_{\infty}(E^{\prime},\omega^{\prime})\oplus
\operatorname{Sp}_{\infty}(E^{\prime\prime},\omega^{\prime\prime})$ is
identified in the obvious way with a subgroup of $\operatorname{Sp}_{\infty
}(E,\omega)$; a similar property holds for the reduced index $m_{\ell}$.

\section{The\ index $\nu$ on $\operatorname{Sp}_{\infty}(n)$}

In this section we define and study a function $\nu:\operatorname*{Sp}%
_{\infty}(n)\longrightarrow\mathbb{Z}$ extending the Conley--Zehnder index
\cite{CZ}. We begin by recalling the definition and main properties of the latter.

\subsection{Review of the Conley--Zehnder index\label{defcz}}

Let $\Sigma$ be a continuous path $[0,1]\longrightarrow\operatorname*{Sp}(n)$
such that $\Sigma(0)=I$ and $\det(\Sigma(1)-I)\neq0$. Loosely speaking, the
Conley--Zehnder index \cite{CZ} counts algebraically the number of points in
the open interval $]0,1[$ for which $\Sigma(t)$ has $1$ as an eigenvalue. To
give a more precise definition we need some notations. Let us define three
subsets of $\operatorname{Sp}(n)$ by%
\begin{align*}
\operatorname{Sp}_{0}(n)  &  =\{S:\det(S-I)=0\}\\
\operatorname{Sp}^{+}(n)  &  =\{S:\det(S-I)>0\}\\
\operatorname{Sp}^{-}(n)  &  =\{S:\det(S-I)<0\}\text{.}%
\end{align*}
These sets partition $\operatorname{Sp}(n)$, and $\operatorname{Sp}^{+}(n)$
and $\operatorname{Sp}^{-}(n)$ are moreover arcwise connected; the symplectic
matrices $S^{+}=-I$ and
\[
S^{-}=%
\begin{bmatrix}
L & 0\\
0 & L^{-1}%
\end{bmatrix}
\text{ \ , \ }L=\operatorname*{diag}[2,-1,...,-1]
\]
belong to $\operatorname{Sp}^{+}(n)$ and $\operatorname{Sp}^{-}(n)$, respectively.

Let us now denote by $C_{\pm}(2n,\mathbb{R})$ the space of all paths
$\Sigma:[0,1]\longrightarrow\operatorname*{Sp}(n)$ with $\Sigma(0)=I$ and
$\Sigma(1)\in\operatorname{Sp}^{\pm}(n)$. Any such path can be extended into a
path $\widetilde{\Sigma}:[0,2]\longrightarrow\operatorname*{Sp}(n)$ such that
$\widetilde{\Sigma}(t)\in\operatorname{Sp}^{\pm}(n)$ for $1\leq t\leq2$ and
$\widetilde{\Sigma}(2)=S^{+}$ or $\widetilde{\Sigma}(2)=S^{-}$. Let $\rho$ be
the mapping $\operatorname*{Sp}(n)\longrightarrow S^{1}$, $\rho(S_{t})=\det
u_{t}$, used in the definition of the Maslov index for symplectic loops. The
Conley--Zehnder index of $\Sigma$ is, by definition, the winding number of the
loop $(\rho\circ\tilde{\Sigma})^{2}$ in $S^{1}$:
\[
i_{\text{CZ}}(\Sigma)=\deg[t\longmapsto(\rho(\tilde{\Sigma}(t)))^{2}\text{,
}0\leq t\leq2]\text{.}%
\]
It turns out that $i_{\text{CZ}}(\Sigma)$ is invariant under homotopy as long
as the endpoint $S=\Sigma(1)$ remains in $\operatorname{Sp}^{\pm}(n)$; in
particular it does not change under homotopies with fixed endpoints so we may
view $i_{\text{CZ}}$ as defined on the subset
\[
\operatorname{Sp}_{\infty}^{\ast}(n)=\{S_{\infty}:\det(S-I)\neq0\}
\]
of the universal covering group $\operatorname{Sp}_{\infty}(n)$. With this
convention one proves \cite{HWZ} that the Conley-Zehnder index is the unique
mapping $i_{\text{CZ}}:\operatorname{Sp}_{\infty}^{\ast}(n)\longrightarrow
\mathbb{Z}$ having the following properties:

\begin{description}
\item[(CZ$_{1}$)] \textit{Antisymmetry}: For every $S_{\infty}$ we have
\[
i_{\text{CZ}}(S_{\infty}^{-1})=-i_{\text{CZ}}(S_{\infty})
\]
where $S_{\infty}^{-1}$ is the homotopy class of the path $t\longmapsto
S_{t}^{-1}$;

\item[(CZ$_{2}$)] \textit{Continuity}: Let $\Sigma$ be a symplectic path
representing $S_{\infty}$ and $\Sigma^{\prime}$ a path joining $S$ to an
element $S^{\prime}$ belonging to the same component $\operatorname{Sp}^{\pm
}(n)$ as $S$. Let $S_{\infty}^{\prime}$ be the homotopy class of $\Sigma
\ast\Sigma^{\prime}$. We have
\[
i_{\text{CZ}}(S_{\infty})=i_{\text{CZ}}(S_{\infty}^{\prime})\text{;}%
\]

\item[(CZ$_{3}$)] \textit{Action of} $\pi_{1}[\operatorname{Sp}(n)]$:
\[
i_{\text{CZ}}(\alpha^{r}S_{\infty})=i_{\text{CZ}}(S_{\infty})+2r
\]
for every $r\in\mathbb{Z}$.
\end{description}

We observe that these three properties are characteristic of the
Conley--Zehnder index in the sense that any other function $i_{\text{CZ}%
}^{\prime}:\operatorname{Sp}_{\infty}^{\ast}(n)\longrightarrow\mathbb{Z}$
satisfying then must be identical to $i_{\text{CZ}}$. Set in fact
$\delta=i_{\text{CZ}}-i_{\text{CZ}}^{\prime}$. In view of (CZ$_{3}$) we have
$\delta(\alpha^{r}S_{\infty})=\delta(S_{\infty})$ for all $r\in\mathbb{Z}$
hence $\delta$ is defined on $\operatorname{Sp}^{\ast}(n)=\operatorname{Sp}%
^{+}(n)\cup\operatorname{Sp}^{-}(n)$ so that $\delta(S_{\infty})=\delta(S)$
where $S=S_{1}$, the endpoint of the path $t\longmapsto S_{t}$. Property
(CZ$_{2}$) implies that this function $\operatorname{Sp}^{\ast}%
(n)\longrightarrow\mathbb{Z}$ is constant on both $\operatorname{Sp}^{+}(n)$
and $\operatorname{Sp}^{-}(n)$. We next observe that since $\det S=1$ we have
$\det(S^{-1}-I)=\det(S-I)$ so that $S$ and $S^{-1}$ always belong to the same
set $\operatorname{Sp}^{+}(n)$ or $\operatorname{Sp}^{-}(n)$ if $\det
(S-I)\neq0$. Property (CZ$_{1}$) then implies that $\delta$ must be zero on
both $\operatorname{Sp}^{+}(n)$ or $\operatorname{Sp}^{-}(n)$.\medskip

Two other noteworthy properties of the Conley--Zehnder are:

\begin{description}
\item[(CZ$_{4}$)] \textit{Normalization}: Let $J_{1}$ be the standard
symplectic matrix in $Sp(1)$. If $S_{1}$ is the path $t\longrightarrow e^{\pi
tJ_{1}}$ ($0\leq t\leq1$) joining $I$ to $-I$ in $\operatorname{Sp}(1)$ then
$i_{\text{CZ},1}(S_{1,\infty})=1$ ($i_{\text{CZ},1}$ the Conley--Zehnder index
on $\operatorname{Sp}(1)$);

\item[(CZ$_{5}$)] \textit{Dimensional additivity}: if $S_{1,\infty}%
\in\operatorname{Sp}_{\infty}^{\ast}(n_{1})$, $S_{2,\infty}\in
\operatorname{Sp}_{\infty}^{\ast}(n_{2})$, $n_{1}+n_{2}=n$, then%
\[
i_{\text{CZ}}(S_{1,\infty}\oplus S_{2,\infty})=i_{\text{CZ,}1}(S_{1,\infty
})+i_{\text{CZ,}2}(S_{2,\infty})
\]
where $i_{\text{CZ,}j}$ is the Conley--Zehnder index on $\operatorname{Sp}%
(n_{j})$, $j=1,2$.
\end{description}

\subsection{Symplectic Cayley transform}

Our extension of the index $i_{\text{CZ}}$ requires a notion of Cayley
transform for symplectic matrices. If $S\in\operatorname*{Sp}(n)$,
$\det(S-I)\neq0$, we call the matrix%
\begin{equation}
M_{S}=\frac{1}{2}J(S+I)(S-I)^{-1} \label{cayley}%
\end{equation}
the \textquotedblleft symplectic Cayley transform of $S$\textquotedblright.
Equivalently:%
\begin{equation}
M_{S}=\frac{1}{2}J+J(S-I)^{-1}. \label{cayleybis}%
\end{equation}
It is straightforward to check that $M_{S}$ always is a symmetric matrix:
$M_{S}=M_{S}^{T}$ (it suffices for this to use the fact that $S^{T}%
JS=SJS^{T}=J$).

The symplectic Cayley transform has in addition the following properties,
which are interesting by themselves:

\begin{lemma}
\label{lemcarton}(i) We have
\begin{equation}
(M_{S}+M_{S^{\prime}})^{-1}=-(S^{\prime}-I)(SS^{\prime}-I)^{-1}(S-I)J
\label{mess}%
\end{equation}
and the symplectic Cayley transform of the product $SS^{\prime}$ is (when
defined) given by the formula%
\begin{equation}
M_{SS^{\prime}}=M_{S}+(S^{T}-I)^{-1}J(M_{S}+M_{S^{\prime}})^{-1}%
J(S-I)^{-1}\text{.} \label{mss}%
\end{equation}
(ii) The symplectic Cayley transform of $S$ and $S^{-1}$ are related by%
\begin{equation}
M_{S^{-1}}=-M_{S}\text{.} \label{rathobvious}%
\end{equation}

\end{lemma}

\begin{proof}
(i) We begin by noting that (\ref{cayleybis}) implies that%
\begin{equation}
M_{S}+M_{S^{\prime}}=J(I+(S-I)^{-1}+(S^{\prime}-I)^{-1}) \label{mesplus}%
\end{equation}
hence the identity (\ref{mess}). In fact, writing $SS^{\prime}-I=S(S^{\prime
}-I)+S-I$, we have%
\begin{align*}
(S^{\prime}-I)(SS^{\prime}-I)^{-1}(S-I)  &  =(S^{\prime}-I)(S(S^{\prime
}-I)+S-I)^{-1}(S-I)\\
&  =((S-I)^{-1}S(S^{\prime}-I)(S^{\prime}-I)^{-1}+(S^{\prime}-I)^{-1})^{-1}\\
&  =((S-I)^{-1}S+(S^{\prime}-I)^{-1})\\
&  =I+(S-I)^{-1}+(S^{\prime}-I)^{-1}\text{;}%
\end{align*}
the equality (\ref{mess}) follows in view of (\ref{mesplus}). Let us prove
(\ref{mss}); equivalently%
\begin{equation}
M_{S}+M=M_{SS^{\prime}} \label{msis}%
\end{equation}
where $M$ is the matrix defined by
\[
M=(S^{T}-I)^{-1}J(M_{S}+M_{S^{\prime}})^{-1}J(S-I)^{-1}%
\]
that is, in view of (\ref{mess}),
\[
M=(S^{T}-I)^{-1}J(S^{\prime}-I)(SS^{\prime}-I)^{-1}\text{.}%
\]
Using the obvious relations $S^{T}=-JS^{-1}J$ and $(-S^{-1}+I)^{-1}%
=S(S-I)^{-1}$ we have
\begin{align*}
M  &  =(S^{T}-I)^{-1}J(S^{\prime}-I)(SS^{\prime}-I)^{-1}\\
&  =-J(-S^{-1}+I)^{-1}(S^{\prime}-I)(SS^{\prime}-I)^{-1}\\
&  =-JS(S-I)^{-1}(S^{\prime}-I)(SS^{\prime}-I)^{-1}%
\end{align*}
that is, writing $S=S-I+I$,
\[
M=-J(S^{\prime}-I)(SS^{\prime}-I)^{-1}-J(S-I)^{-1}(S^{\prime}-I)(SS^{\prime
}-I)^{-1}\text{.}%
\]
Replacing $M_{S}$ by its value (\ref{cayleybis}) we have%
\begin{multline*}
M_{S}+M=\\
J(\tfrac{1}{2}I+(S-I)^{-1}-(S^{\prime}-I)(SS^{\prime}-I)^{-1}-(S-I)^{-1}%
(S^{\prime}-I)(SS^{\prime}-I)^{-1})\text{;}%
\end{multline*}
noting that%
\begin{multline*}
(S-I)^{-1}-(S-I)^{-1}(S^{\prime}-I)(SS^{\prime}-I)^{-1}=\\
(S-I)^{-1}(SS^{\prime}-I-S^{\prime}+I)(SS^{\prime}-I)^{-1})
\end{multline*}
that is%
\begin{align*}
(S-I)^{-1}-(S-I)^{-1}(S^{\prime}-I)(SS^{\prime}-I)^{-1}  &  =(S-I)^{-1}%
(SS^{\prime}-S^{\prime})(SS^{\prime}-I)^{-1}\\
&  =S^{\prime}(SS^{\prime}-I)^{-1})
\end{align*}
we get%
\begin{align*}
M_{S}+M  &  =J(\tfrac{1}{2}I-(S^{\prime}-I)(SS^{\prime}-I)^{-1}+S^{\prime
}(SS^{\prime}-I)^{-1})\\
&  =J(\tfrac{1}{2}I+(SS^{\prime}-I)^{-1})\\
&  =M_{SS^{\prime}}%
\end{align*}
which we set out to prove. (ii) Formula (\ref{rathobvious}) follows from the
sequence of equalities%
\begin{align*}
M_{S^{-1}}  &  =\tfrac{1}{2}J+J(S^{-1}-I)^{-1}\\
&  =\tfrac{1}{2}J-JS(S-I)^{-1}\\
&  =\tfrac{1}{2}J-J(S-I+I)(S-I)^{-1}\\
&  =-\tfrac{1}{2}J-J(S-I)^{-1}\\
&  =-M_{S}\text{.}%
\end{align*}

\end{proof}

\subsection{Definition and properties of $\nu(S_{\infty})$}

We define on $\mathbb{R}^{2n}\oplus\mathbb{R}^{2n}$ a symplectic form
$\sigma^{\ominus}$ by%
\[
\sigma^{\ominus}(z_{1},z_{2};z_{1}^{\prime},z_{2}^{\prime})=\sigma(z_{1}%
,z_{1}^{\prime})-\sigma(z_{2},z_{2}^{\prime})
\]
and denote by $\operatorname*{Sp}^{\ominus}(2n)$ and $\operatorname{Lag}%
^{\ominus}(2n)$ the corresponding symplectic group and Lagrangian
Grassmannian. Let $\mu^{\ominus}$ be the ALM index on $\operatorname{Lag}%
_{\infty}^{\ominus}(2n)$ and $\mu_{L}^{\ominus}$ the Maslov index on
$\operatorname*{Sp}_{\infty}^{\ominus}(2n)$ relative to $L\in
\operatorname{Lag}^{\ominus}(2n)$.

For $S_{\infty}\in\operatorname*{Sp}_{\infty}(n)$ we define
\begin{equation}
\nu(S_{\infty})=\frac{1}{2}\mu^{\ominus}((I\oplus S)_{\infty}\Delta_{\infty
},\Delta_{\infty}) \label{rscz}%
\end{equation}
where $(I\oplus S)_{\infty}$ is the homotopy class in $\operatorname{Sp}%
^{\ominus}(2n)$ of the path
\[
t\longmapsto\{(z,S_{t}z):z\in\mathbb{R}^{2n}\}\text{ \ , \ }0\leq t\leq1
\]
and $\Delta=\{(z,z):z\in\mathbb{R}^{2n}\}$ the diagonal of $\mathbb{R}%
^{2n}\oplus\mathbb{R}^{2n}$. Setting $S_{t}^{\ominus}=I\oplus S_{t}$ we have
$S_{t}^{\ominus}\in\operatorname*{Sp}^{\ominus}(2n)$ hence formulae
(\ref{rscz}) is equivalent to
\begin{equation}
\nu(S_{\infty})=\frac{1}{2}\mu_{\Delta}^{\ominus}(S_{\infty}^{\ominus})
\label{muis}%
\end{equation}
where $\mu_{\Delta}^{\ominus}$ is the relative Maslov index on
$\operatorname*{Sp}_{\infty}^{\ominus}(2n)$ corresponding to the choice
$\Delta\in\operatorname{Lag}^{\ominus}(2n)$.

Note that replacing $n$ by $2n$ in the congruence (\ref{mod2}) we have
\begin{align*}
\mu^{\ominus}((I\oplus S)_{\infty}\Delta_{\infty},\Delta_{\infty})  &
\equiv\dim((I\oplus S)\Delta\cap\Delta)\text{ \ }\operatorname{mod}2\\
&  \equiv\dim\operatorname{Ker}(S-I)\text{ \ }\operatorname{mod}2
\end{align*}
and hence%
\[
\nu(S_{\infty})\equiv\frac{1}{2}\dim\operatorname{Ker}(S-I)\text{
\ }\operatorname{mod}1\text{.}%
\]
Since the eigenvalue $1$ of $S$ has even multiplicity $\nu(S_{\infty})$ is
thus always an integer.

The index $\nu$ has the following three important properties; the third is
essential for the calculation of the index of repeated periodic orbits (it
clearly shows that $\nu$ is not in general additive):

\begin{proposition}
\label{propbasic}(i) For all $S_{\infty}\in\operatorname*{Sp}_{\infty}(n)$ we
have%
\begin{equation}
\nu(S_{\infty}^{-1})=-\nu(S_{\infty})\text{ \ , \ \ }\nu(I_{\infty})=0
\label{antinu}%
\end{equation}
($I_{\infty}$ the identity of the group $\operatorname*{Sp}_{\infty}(n)$).
(ii) For all $r\in\mathbb{Z}$ we have
\begin{equation}
\nu(\alpha^{r}S_{\infty})=\nu(S_{\infty})+2r\text{ \ , \ }\nu(\alpha^{r})=2r
\label{anu}%
\end{equation}
(iii) Let $S_{\infty}$ be the homotopy class of a path $\Sigma$ in
$\operatorname*{Sp}(n)$ joining the identity to $S\in\operatorname*{Sp}%
\nolimits^{\ast}(n)$, and let $S^{\prime}\in\operatorname*{Sp}(n)$ be in the
same connected component $\operatorname*{Sp}\nolimits^{\pm}(n)$ as $S$. Then
$\nu(S_{\infty}^{\prime})=\nu(S_{\infty})$ where $S_{\infty}^{\prime}$ is the
homotopy class in $\operatorname*{Sp}(n)$ of the concatenation of $\Sigma$ and
a path joining $S$ to $S^{\prime}$ in $\operatorname*{Sp}\nolimits_{0}(n)$.
\end{proposition}

\begin{proof}
(i) Formulae (\ref{antinu}) immediately follows from the equality $(S_{\infty
}^{\ominus})^{-1}=(I\oplus S^{-1})_{\infty}$ and the antisymmetry of
$\mu_{\Delta}^{\ominus}$. (ii) The second formula (\ref{anu}) follows from the
first using (\ref{antinu}). To prove the first formula (\ref{anu}) it suffices
to observe that to the generator $\alpha$ of $\pi_{1}[\operatorname{Sp}(n)]$
corresponds the generator $I_{\infty}\oplus\alpha$ of $\pi_{1}%
[\operatorname*{Sp}^{\ominus}(2n)]$; in view of property (\ref{sex}) of the
Maslov index it follows that
\begin{align*}
\nu(\alpha^{r}S_{\infty})  &  =\frac{1}{2}\mu_{\Delta}^{\ominus}((I_{\infty
}\oplus\alpha)^{r}S_{\infty}^{\ominus})\\
&  =\frac{1}{2}(\mu_{\Delta}^{\ominus}(S_{\infty}^{\ominus})+4r)\\
&  =\nu(S_{\infty})+2r\text{.}%
\end{align*}
(iii) Assume in fact that $S$ and $S^{\prime}$ belong to, say,
$\operatorname{Sp}^{+}(n)$. Let $S_{\infty}$ be the homotopy class of the path
$\Sigma$, and $\Sigma^{\prime}$ a path joining $S$ to $S^{\prime}$ in
$\operatorname{Sp}^{+}(n)$ (we parametrize both paths by $t\in\lbrack0,1]$).
Let $\Sigma_{t^{\prime}}^{\prime}$ be the restriction of $\Sigma^{\prime}$ to
the interval $[0,t^{\prime}]$, $t^{\prime}\leq t$ and $S_{\infty}(t^{\prime})$
the homotopy class of the concatenation $\Sigma\ast\Sigma_{t^{\prime}}%
^{\prime}$. We have $\det(S(t)-I)>0$ for all $t\in\lbrack0,t^{\prime}]$ hence
$S_{\infty}^{\ominus}(t)\Delta\cap\Delta\neq0$ as $t$ varies from $0$ to $1$.
It follows from the fact that the $\mu_{\Delta}^{\ominus}$ is locally constant
on $\{S_{\infty}^{\ominus}:S_{\infty}^{\ominus}\Delta\cap\Delta=0\}$ (see
\S \ref{relma}) that the function $t\longmapsto\mu_{\Delta}^{\ominus
}(S_{\infty}^{\ominus}(t))$ is constant, and hence%
\[
\mu_{\Delta}^{\ominus}(S_{\infty}^{\ominus})=\mu_{\Delta}^{\ominus}(S_{\infty
}^{\ominus}(0))=\mu_{\Delta}^{\ominus}(S_{\infty}^{\ominus}(1))=\mu_{\Delta
}^{\ominus}(S_{\infty}^{\prime\ominus})
\]
which was to be proven.\bigskip
\end{proof}

The following consequence of the result above shows that the indices $\nu$ and
$i_{\text{CZ}}$ coincide on their common domain of definition:

\begin{corollary}
The restriction of the index $\nu$ to $\operatorname{Sp}^{\ast}(n)$ is the
Conley--Zehnder index:
\[
\nu(S_{\infty})=i_{\text{CZ}}(S_{\infty})\text{ \ if\ \ }\det(S-I)\neq0.
\]

\end{corollary}

\begin{proof}
The restriction of $\nu$ to $\operatorname{Sp}^{\ast}(n)$ satisfies the
properties (CZ$_{1}$), (CZ$_{2}$), and (CZ$_{3}$) of the Conley--Zehnder index
listed in \S \ref{defcz}; we showed that these properties uniquely
characterize $i_{\text{CZ}}$.\bigskip
\end{proof}

Let us prove a formula for the index of the product of two paths:

\begin{proposition}
If $S_{\infty}$, $S_{\infty}^{\prime}$, and $S_{\infty}S_{\infty}^{\prime}$
are such that $\det(S-I)\neq0$, $\det(S^{\prime}-I)\neq0$, and $\det
(SS^{\prime}-I)\neq0$ then%
\begin{equation}
\nu(S_{\infty}S_{\infty}^{\prime})=\nu(S_{\infty})+\nu(S_{\infty}^{\prime
})+\tfrac{1}{2}\operatorname*{sign}(M_{S}+M_{S^{\prime}}) \label{muss}%
\end{equation}
where $M_{S}$ is the symplectic Cayley transform of $S$; in particular%
\begin{equation}
\nu(S_{\infty}^{r})=r\nu(S_{\infty})+\tfrac{1}{2}(r-1)\operatorname*{sign}%
M_{S} \label{mussrepet}%
\end{equation}
for every integer $r$.
\end{proposition}

\begin{proof}
In view of (\ref{muis}) and the product property (\ref{uno}) of the Maslov
index we have%
\begin{align*}
\nu(S_{\infty}S_{\infty}^{\prime})  &  =\nu(S_{\infty})+\nu(S_{\infty}%
^{\prime})+\tfrac{1}{2}\tau^{\ominus}(\Delta,S^{\ominus}\Delta,S^{\ominus
}S^{\prime\ominus}\Delta)\\
&  =\nu(S_{\infty})+\nu(S_{\infty}^{\prime})-\tfrac{1}{2}\tau^{\ominus
}(S^{\ominus}S^{\prime\ominus}\Delta,S^{\ominus}\Delta,\Delta)
\end{align*}
where $S^{\ominus}=I\oplus S$, $S^{\prime\ominus}=I\oplus S^{\prime}$ and
$\tau^{\ominus}$ is the signature on the symplectic space $(\mathbb{R}%
^{2n}\oplus\mathbb{R}^{2n},\sigma^{\ominus})$. The condition $\det(SS^{\prime
}-I)\neq0$ is equivalent to $S^{\ominus}S^{\prime\ominus}\Delta\cap\Delta=0$
hence we can apply property \textit{(i)} in Lemma \ref{lemsimple} with
$\ell=S^{\ominus}S^{\prime\ominus}\Delta$, $\ell^{\prime}=S^{\ominus}\Delta$,
and $\ell^{\prime\prime}=\Delta$. The projection operator onto $S^{\ominus
}S^{\prime\ominus}\Delta$ along $\Delta$ is easily seen to be%
\[
\Pr\nolimits_{S^{\ominus}S^{\prime\ominus}\Delta,\Delta}=%
\begin{bmatrix}
(I-SS^{\prime})^{-1} & -(I-SS^{\prime})^{-1}\smallskip\\
SS^{\prime}(I-SS^{\prime})^{-1} & -SS^{\prime}(I-SS^{\prime})^{-1}%
\end{bmatrix}
\]
hence $\tau^{\ominus}(S^{\ominus}S^{\prime\ominus}\Delta,S^{\ominus}%
\Delta,\Delta)$ is the signature of the quadratic form%
\[
Q(z)=\sigma^{\ominus}(\Pr\nolimits_{S^{\ominus}S^{\prime\ominus}\Delta,\Delta
}(z,Sz);(z,Sz))
\]
that is, since $\sigma^{\ominus}=\sigma\ominus\sigma$:%
\begin{align*}
Q(z)  &  =\sigma((I-SS^{\prime})^{-1}(I-S)z,z))-\sigma(SS^{\prime
}(I-SS^{\prime})^{-1}(I-S)z,Sz))\\
&  =\sigma((I-SS^{\prime})^{-1}(I-S)z,z))-\sigma(S^{\prime}(I-SS^{\prime
})^{-1}(I-S)z,z))\\
&  =\sigma((I-S^{\prime})(I-SS^{\prime})^{-1}(I-S)z,z))\text{.}%
\end{align*}
In view of formula (\ref{mess}) in Lemma \ref{lemcarton} we have%
\[
(I-S^{\prime})(SS^{\prime}-I)^{-1}(I-S)=(M_{S}+M_{S^{\prime}})^{-1}J
\]
hence
\[
Q(z)=-\left\langle (M_{S}+M_{S^{\prime}})^{-1}Jz,Jz\right\rangle
\]
and the signature of $Q$ is thus the same as that of%
\[
Q^{\prime}(z)=-\left\langle (M_{S}+M_{S^{\prime}})^{-1}z,z\right\rangle
\]
that is $-\operatorname*{sign}(M_{S}+M_{S^{\prime}})$. This proves formula
(\ref{muss}). Formula (\ref{mussrepet}) follows from (\ref{muss}) by induction
on $r$.\bigskip
\end{proof}

It is often deplored in the literature on Gutzwiller's formula (\ref{Gutz})
that it is not always obvious that the index $\mu_{\gamma}$ of the periodic
orbit $\gamma$ is independent on the choice of the origin of the orbit. Let us
prove that this property always holds:

\begin{proposition}
Let $(f_{t})$ be the flow determined by a (time-independent) Hamiltonian
function on $\mathbb{R}^{2n}$ and $z\neq0$ such that $f_{T}(z)=z$ for some
$T>0$. Let $z^{\prime}=f_{t^{\prime}}(z)$ for some $t^{\prime}$ and denote by
$S_{T}(z)=Df_{T}(z)$ and $S_{T}(z^{\prime})=Df_{T}(z^{\prime})$ the
corresponding monodromy matrices. Let $S_{T}(z)_{\infty}$ and $S_{T}%
(z^{\prime})_{\infty}$ be the homotopy classes of the paths $t\longmapsto
S_{t}(z)=Df_{t}(z)$ and $t\longmapsto S_{t}(z^{\prime})=Df_{t}(z^{\prime})$,
$0\leq t\leq T$. We have $\nu(S_{T}(z)_{\infty})=\nu(S_{T}(z^{\prime}%
)_{\infty})$.
\end{proposition}

\begin{proof}
The monodromy matrices $S_{T}(z)$ and $S_{T}(z^{\prime})$ are conjugate of
each other; in fact (proof of Theorem 6 in \cite{MSDG}):
\[
S_{T}(z^{\prime})=S_{t^{\prime}}(z^{\prime})S_{T}(z)S_{t^{\prime}}(z^{\prime
})^{-1}\text{;}%
\]
since we will need to let $t^{\prime}$ vary we write $S_{T}(z^{\prime}%
)=S_{T}(z^{\prime},t^{\prime})$ so that%
\[
S_{T}(z^{\prime},t^{\prime})=S_{t^{\prime}}(z^{\prime})S_{T}(z)S_{t^{\prime}%
}(z^{\prime})^{-1}\text{.}%
\]
The paths $t\longmapsto S_{t}(z^{\prime})$ and $t\longmapsto S_{t^{\prime}%
}(z^{\prime})S_{t}(z)S_{t^{\prime}}(z^{\prime})^{-1}$ being homotopic with
fixed endpoints $S_{T}(z^{\prime},t^{\prime})_{\infty}$ is also the homotopy
class of the path $t\longmapsto S_{t^{\prime}}(z^{\prime})S_{t}(z)S_{t^{\prime
}}(z^{\prime})^{-1}$. We thus have, by definition (\ref{rscz}) of $\nu$,
\[
\nu(S_{T}(z^{\prime},t^{\prime})_{\infty})=\frac{1}{2}\mu_{\Delta_{t^{\prime}%
}}^{\ominus}(S_{T}^{\ominus}(z)_{\infty})
\]
where we have set%
\[
\Delta_{t^{\prime}}=(I\oplus S_{t^{\prime}}(z^{\prime})^{-1})\Delta\text{
\ and \ }S_{T}^{\ominus}(z)_{\infty}=I_{\infty}\oplus S_{T}(z)_{\infty
}\text{.}%
\]
Consider now the mapping $t^{\prime}\longmapsto\mu_{\Delta_{t^{\prime}}%
}^{\ominus}(S_{T}^{\ominus}(z)_{\infty})$; we have%
\[
S_{T}^{\ominus}(z)\Delta_{t^{\prime}}\cap\Delta_{t^{\prime}}=\{z:Sz=z\}
\]
hence the dimension of the intersection $S_{T}^{\ominus}(z)\Delta_{t^{\prime}%
}\cap\Delta_{t^{\prime}}$ remains constant as $t^{\prime}$ varies; in view of
the topological property of the relative Maslov index the mapping $t^{\prime
}\longmapsto\mu_{\Delta_{t^{\prime}}}^{\ominus}(S_{T}^{\ominus}(z)_{\infty})$
is thus constant and hence%
\[
\nu(S_{T}(z^{\prime},t^{\prime})_{\infty})=\nu(S_{T}(z^{\prime},0)_{\infty
})=\nu(S_{T}(z)_{\infty})
\]
which concludes the proof.
\end{proof}

\subsection{Relation between $\nu$ and $\mu_{\ell_{P}}$}

The index $\nu$ can be expressed in simple -- and useful -- way in terms of
the Maslov index $\mu_{\ell_{P}}$ on $\operatorname*{Sp}_{\infty}(n)$. The
following technical result will be helpful in establishing this relation.
Recall that $S\in\operatorname*{Sp}(n)$ is said to be \textquotedblleft
free\textquotedblright\ if we have $S\ell_{P}\cap\ell_{P}=0$; this condition
is equivalent to $\det B\neq0$ when $S$ is identified with the matrix
\begin{equation}
S=%
\begin{bmatrix}
A & B\\
C & D
\end{bmatrix}
\label{czfree}%
\end{equation}
in the canonical basis. The set of all free symplectic matrices is dense in
$\operatorname*{Sp}(n)$. The quadratic form $W$ on $\mathbb{R}_{x}^{n}%
\times\mathbb{R}_{x}^{n}$ defined by
\[
W(x,x^{\prime})=\frac{1}{2}\left\langle Px,x\right\rangle -\left\langle
Lx,x^{\prime}\right\rangle +\frac{1}{2}\left\langle Qx^{\prime},x^{\prime
}\right\rangle
\]
where
\begin{equation}
P=DB^{-1}\text{, }L=B^{-1}\text{, }Q=B^{-1}A \label{pabcd}%
\end{equation}
then generates $S$ in the sense that
\[
(x,p)=S(x^{\prime},p^{\prime})\Longleftrightarrow p=\partial_{x}W(x,x^{\prime
})\text{ , }p^{\prime}=-\partial_{x^{\prime}}W(x,x^{\prime}).
\]

We have:

\begin{lemma}
\label{lemma1}Let $S=S_{W}\in\operatorname*{Sp}(n)$ be given by (\ref{czfree}%
).We have
\begin{equation}
\det(S_{W}-I)=(-1)^{n}\det B\det(B^{-1}A+DB^{-1}-B^{-1}-(B^{T})^{-1})
\label{bofor1}%
\end{equation}
that is:%
\[
\det(S_{W}-I)=(-1)^{n}\det(L^{-1})\det(P+Q-L-L^{T})\text{.}%
\]
In particular the symmetric matrix
\[
P+Q-L-L^{T}=DB^{-1}+B^{-1}A-B^{-1}-(B^{T})^{-1}%
\]
is invertible.
\end{lemma}

\begin{proof}
Since $B$ is invertible we can factorize $S-I$ as%
\[%
\begin{bmatrix}
A-I & B\\
C & D-I
\end{bmatrix}
=%
\begin{bmatrix}
0 & B\\
I & D-I
\end{bmatrix}%
\begin{bmatrix}
C-(D-I)B^{-1}(A-I) & 0\\
B^{-1}(A-I) & I
\end{bmatrix}
\]
and hence%
\begin{align*}
\det(S_{W}-I)  &  =\det(-B)\det(C-(D-I)B^{-1}(A-I))\\
&  =(-1)^{n}\det B\det(C-(D-I)B^{-1}(A-I))\text{.}%
\end{align*}
Since $S$ is symplectic we have $C-DB^{-1}A=-(B^{T})^{-1}$ and hence%
\[
C-(D-I)B^{-1}(A-I))=B^{-1}A+DB^{-1}-B^{-1}-(B^{T})^{-1}\text{;}%
\]
the Lemma follows.\bigskip
\end{proof}

Let us now introduce the notion of index of concavity of a Hamiltonian
periodic orbit $\gamma$, defined for $0\leq t\leq T$, with $\gamma
(0)=\gamma(T)=z_{0}$. As $t$ goes from $0$ to $T$ the linearized part
$D\gamma(t)=S_{t}(z_{0})$ goes from the identity to $S_{T}(z_{0})$ (the
monodromy matrix) in $\operatorname{Sp}(n)$. We assume that $S_{T}(z_{0})$ is
free and that $\det(S_{T}(z_{0})-I)\neq0$. Writing%
\[
S_{t}(z_{0})=%
\begin{bmatrix}
A(t) & B(t)\\
C(t) & D(t)
\end{bmatrix}
\]
we thus have $\det B(t)\neq0$ in a neighborhood $[T-\varepsilon,T+\varepsilon
]$ of the time $T$. The generating function
\[
W(x,x^{\prime},t)=\frac{1}{2}\left\langle P(t)x,x\right\rangle -\left\langle
L(t)x,x^{\prime}\right\rangle +\frac{1}{2}\left\langle Q(t)x^{\prime
},x^{\prime}\right\rangle
\]
(with $P(t)$, $Q(t)$, $L(t)$ defined by (\ref{pabcd}) thus exists for
$T-\varepsilon\leq t\leq T+\varepsilon$. By definition Morse's index of
concavity \cite{Morse} (also see \cite{MG,Piccione}) of the periodic orbit
$\gamma$ is the index of inertia
\[
\operatorname*{Inert}W_{xx}^{\prime\prime}=\operatorname*{Inert}(P+Q-L-L^{T})
\]
of $W_{xx}^{\prime\prime}$, the matrix of second derivatives of the function
$x\longmapsto W(x,x;T)$ (we have set $P=P(T)$, $Q=Q(T)$, $L=L(T)$).

Let us now prove the following essential result; recall that $m_{\ell}$
denotes the reduced Maslov index (\ref{remas}) associated to $\mu_{\ell}$:

\begin{proposition}
\label{propconca}Let $t\longmapsto S_{t}$ be a symplectic path, $0\leq t\leq
1$. Let $S_{\infty}\in\operatorname*{Sp}_{\infty}(n)$ be the homotopy class of
that path and set $S=S_{1}$. If $\det(S-I)\neq0$ and $S\ell_{P}\cap\ell_{P}=0$
then%
\begin{equation}
\nu(S_{\infty})=\frac{1}{2}(\mu_{\ell_{P}}(S_{\infty})+\operatorname*{sign}%
W_{xx}^{\prime\prime})=m_{\ell_{P}}(S_{\infty})-\operatorname*{Inert}%
W_{xx}^{\prime\prime} \label{muw}%
\end{equation}
where $\operatorname*{Inert}W_{xx}^{\prime\prime}$ is the index of concavity
corresponding to the endpoint $S$ of the path $t\longmapsto S_{t}$.
\end{proposition}

\begin{proof}
We will divide the proof in three steps. \textit{Step 1.} Let $L\in
\operatorname{Lag}^{\ominus}(4n)$. Using successively formulae (\ref{muis})
and (\ref{sl}) we have%
\begin{equation}
\nu(S_{\infty})=\frac{1}{2}(\mu_{L}^{\ominus}(S_{\infty}^{\ominus}%
)+\tau^{\ominus}(S^{\ominus}\Delta,\Delta,L)-\tau^{\ominus}(S^{\ominus}%
\Delta,S^{\ominus}L,L))\text{.} \label{star}%
\end{equation}
Choosing in particular $L=L_{0}=\ell_{P}\oplus\ell_{P}$ we get
\begin{align*}
\mu_{L_{0}}^{\ominus}(S_{\infty}^{\ominus})  &  =\mu^{\ominus}((I\oplus
S)_{\infty}(\ell_{P}\oplus\ell_{P}),(\ell_{P}\oplus\ell_{P}))\\
&  =\mu(\ell_{P,\infty},\ell_{P,\infty})-\mu(\ell_{P,\infty},S_{\infty}%
\ell_{P,\infty})\\
&  =-\mu(\ell_{P,\infty},S_{\infty}\ell_{P,\infty})\\
&  =\mu_{\ell_{P}}(S_{\infty})
\end{align*}
so that there remains to prove that
\[
\tau^{\ominus}(S^{\ominus}\Delta,\Delta,L_{0})-\tau^{\ominus}(S^{\ominus
}\Delta,S^{\ominus}L_{0},L_{0})=-2\operatorname*{sign}W_{xx}^{\prime\prime
}\text{.}%
\]
\textit{Step 2.} We are going to show that%
\[
\tau^{\ominus}(S^{\ominus}\Delta,S^{\ominus}L_{0},L_{0})=0\text{;}%
\]
in view of the symplectic invariance and the antisymmetry of $\tau^{\ominus}$
this is equivalent to%
\begin{equation}
\tau^{\ominus}(L_{0},\Delta,L_{0},(S^{\ominus})^{-1}L_{0})=0\text{.}
\label{cucu}%
\end{equation}
We have
\[
\Delta\cap L_{0}=\{(0,p;0,p):p\in\mathbb{R}^{n}\}
\]
and $(S^{\ominus})^{-1}L_{0}\cap L_{0}$ consists of all $(0,p^{\prime}%
,S^{-1}(0,p^{\prime\prime}))$ with $S^{-1}(0,p^{\prime\prime})=(0,p^{\prime}%
)$; since $S$ (and hence $S^{-1}$) is free we must have $p^{\prime}%
=p^{\prime\prime}=0$ so that%
\[
(S^{\ominus})^{-1}L_{0}\cap L_{0}=\{(0,p;0,0):p\in\mathbb{R}^{n}\}\text{.}%
\]
It follows that we have%
\[
L_{0}=\Delta\cap L_{0}+(S^{\ominus})^{-1}L_{0}\cap L_{0}%
\]
hence (\ref{cucu}) in view of property \textit{(ii)} in Lemma \ref{lemsimple}.
\textit{Step 3}. Let us finally prove that.%
\[
\tau^{\ominus}(S^{\ominus}\Delta,\Delta,L_{0})=-2\operatorname*{sign}%
W_{xx}^{\prime\prime}\text{;}%
\]
this will complete the proof of the proposition. The condition $\det
(S-I)\neq0$ is equivalent to $S^{\ominus}\Delta\cap\Delta=0$ hence, using
property \textit{(i)} in Lemma \ref{lemsimple}:%
\[
\tau^{\ominus}(S^{\ominus}\Delta,\Delta,L_{0})=-\tau^{\ominus}(S^{\ominus
}\Delta,L_{0},\Delta)
\]
is the signature of the quadratic form $Q$ on $L_{0}$ defined by
\[
Q(0,p,0,p^{\prime})=-\sigma^{\ominus}(\Pr\nolimits_{S^{\ominus}\Delta,\Delta
}(0,p,0,p^{\prime});0,p,0,p^{\prime})
\]
where
\[
\Pr\nolimits_{S^{\ominus}\Delta,\Delta}=%
\begin{bmatrix}
(S-I)^{-1} & -(S-I)^{-1}\smallskip\\
S(S-I)^{-1} & -S(S-I)^{-1}%
\end{bmatrix}
\]
is the projection on $S^{\ominus}\Delta$ along $\Delta$ in $\mathbb{R}%
^{2n}\oplus\mathbb{R}^{2n}$. It follows that the quadratic form $Q$ is given
by%
\[
Q(0,p,0,p^{\prime})=-\sigma^{\ominus}((I-S)^{-1}(0,p^{\prime\prime
}),S(I-S)^{-1}(0,p^{\prime\prime});0,p,0,p^{\prime})
\]
where we have set $p^{\prime\prime}=p-p^{\prime}$; by definition of
$\sigma^{\ominus}$ this is%
\[
Q(0,p,0,p^{\prime})=-\sigma((I-S)^{-1}(0,p^{\prime\prime}),(0,p))+\sigma
(S(I-S)^{-1}(0,p^{\prime\prime}),(0,p^{\prime}))\text{. }%
\]
Let now $M_{S}$ be the symplectic Cayley transform (\ref{cayley}) of $S$; we
have%
\[
(I-S)^{-1}=JM_{S}+\tfrac{1}{2}I\text{ \ , \ }S(I-S)^{-1}=JM_{S}-\tfrac{1}{2}I
\]
and hence%
\begin{align*}
Q(0,p,0,p^{\prime})  &  =-\sigma((JM_{S}+\tfrac{1}{2}I)(0,p^{\prime\prime
}),(0,p))+\sigma((JM_{S}-\tfrac{1}{2}I)(0,p^{\prime\prime}),(0,p^{\prime}))\\
&  =-\sigma(JM_{S}(0,p^{\prime\prime}),(0,p))+\sigma(JM_{S}(0,p^{\prime\prime
}),(0,p^{\prime}))\\
&  =\sigma(JM_{S}(0,p^{\prime\prime}),(0,p^{\prime\prime}))\\
&  =-\left\langle M_{S}(0,p^{\prime\prime}),(0,p^{\prime\prime})\right\rangle
\text{.}%
\end{align*}
Let us calculate explicitly $M_{S}$. Writing $S$ in usual block-form we have%
\[
S-I=%
\begin{bmatrix}
0 & B\smallskip\\
I & D-I
\end{bmatrix}%
\begin{bmatrix}
C-(D-I)B^{-1}(A-I) & 0\smallskip\\
B^{-1}(A-I) & I
\end{bmatrix}
\]
that is%
\[
S-I=%
\begin{bmatrix}
0 & B\\
I & D-I
\end{bmatrix}%
\begin{bmatrix}
W_{xx}^{\prime\prime} & 0\\
B^{-1}(A-I) & I
\end{bmatrix}
\]
where we have used the identity
\[
C-(D-I)B^{-1}(A-I))=B^{-1}A+DB^{-1}-B^{-1}-(B^{T})^{-1}%
\]
which follows from the relation $C-DB^{-1}A=-(B^{T})^{-1}$ (the latter is a
rephrasing of the equalities $D^{T}A-B^{T}C=I$ and $D^{T}B=B^{T}D$, which
follow from the fact that $S^{T}JS=S^{T}JS$ since $S\in\operatorname*{Sp}%
(n)$). It follows that%
\begin{align*}
(S-I)^{-1}  &  =%
\begin{bmatrix}
(W_{xx}^{\prime\prime})^{-1} & 0\smallskip\\
B^{-1}(I-A)(W_{xx}^{\prime\prime})^{-1} & I
\end{bmatrix}%
\begin{bmatrix}
(I-D)B^{-1} & I\smallskip\\
B^{-1} & 0
\end{bmatrix}
\medskip\\
&  =%
\begin{bmatrix}
(W_{xx}^{\prime\prime})^{-1}(I-D)B^{-1} & (W_{xx}^{\prime\prime}%
)^{-1}\smallskip\\
B^{-1}(I-A)(W_{xx}^{\prime\prime})^{-1}(I-D)B^{-1}+B^{-1} & B^{-1}%
(I-A)(W_{xx}^{\prime\prime})^{-1}%
\end{bmatrix}
\end{align*}
and hence%
\[
M_{S}=%
\begin{bmatrix}
B^{-1}(I-A)(W_{xx}^{\prime\prime})^{-1}(I-D)B^{-1}+B^{-1} & \frac{1}%
{2}I+B^{-1}(I-A)(W_{xx}^{\prime\prime})^{-1}\smallskip\\
-\frac{1}{2}I-(W_{xx}^{\prime\prime})^{-1}(I-D)B^{-1} & -(W_{xx}^{\prime
\prime})^{-1}%
\end{bmatrix}
\]
from which follows that%
\begin{align*}
Q(0,p,0,p^{\prime})  &  =\left\langle (W_{xx}^{\prime\prime})^{-1}%
p^{\prime\prime},p^{\prime\prime}\right\rangle \\
&  =\left\langle (W_{xx}^{\prime\prime})^{-1}(p-p^{\prime}),(p-p^{\prime
})\right\rangle \text{.}%
\end{align*}
The matrix of the quadratic form $Q$ is thus%
\[
2%
\begin{bmatrix}
(W_{xx}^{\prime\prime})^{-1} & -(W_{xx}^{\prime\prime})^{-1}\smallskip\\
-(W_{xx}^{\prime\prime})^{-1} & (W_{xx}^{\prime\prime})^{-1}%
\end{bmatrix}
\]
and this matrix has signature $2\operatorname*{sign}(W_{xx}^{\prime\prime
})^{-1}=2\operatorname*{sign}W_{xx}^{\prime\prime}$, proving the first
equality (\ref{muw}); the second equality follows because $\mu_{\ell_{P}%
}(S_{\infty})=2m_{\ell_{P}}(S_{\infty})-n$ since $S\ell_{P}\cap\ell_{P}=0$ and
the fact that $W_{xx}^{\prime\prime}$ has rank $n$ in view of Lemma
\ref{lemma1}.\bigskip
\end{proof}

\begin{remark}
Lemma \ref{lemma1} above shows that if $S$ is free then we have
\begin{align*}
\frac{1}{\pi}\arg\det(S-I)  &  \equiv n+\arg\det B+\arg\det W_{xx}%
^{\prime\prime}\text{ \ }\operatorname{mod}2\\
&  \equiv n-\arg\det B+\arg\det W_{xx}^{\prime\prime}\text{ \ }%
\operatorname{mod}2
\end{align*}
In \cite{AIF,Cocycles} we have shown that the reduced Maslov index
$m_{\ell_{P}}(S_{\infty})$ corresponds to a choice of $\arg\det B$ modulo $4$;
Proposition \ref{propconca} thus justifies the following definition of the
argument of $\det(S-I)$:%
\[
\frac{1}{\pi}\arg\det(S-I)\equiv n-\nu(S_{\infty})\text{ \ }\operatorname{mod}%
4.
\]
That this is indeed the correct choice modulo $4$ has been proven by other
means (the Weyl theory of the metaplectic group) by one of us in a recent
publication \cite{mdglettmath}.
\end{remark}

\section{An Example}

Let us begin with a very simple situation. Consider the one-dimensional
harmonic oscillator with Hamiltonian function%
\[
H=\frac{\omega}{2}(p^{2}+x^{2})\text{;}%
\]
all the orbits are periodic with period $2\pi/\omega$. The monodromy matrix is
simply the identity: $\Sigma_{T}=I$ where%
\[
\Sigma_{t}=%
\begin{bmatrix}
\cos\omega t & \sin\omega t\\
-\sin\omega t & \cos\omega t
\end{bmatrix}
\text{.}%
\]
Let us calculate the corresponding index $\nu(\Sigma_{\infty})$. The homotopy
class of path $t\longmapsto\Sigma_{t}$ as $t$ goes from $0$ to $T=2\pi/\omega$
is just the inverse of $\alpha$, the generator of $\pi_{1}[\operatorname{Sp}%
(1)]$ hence $\nu(\Sigma_{\infty})=-2$ in view of (\ref{anu}). If we had
considered $r$ repetitions of the orbit we would likewise have obtained
$\nu(\Sigma_{\infty})=-2r$.

Consider next a two-dimensional harmonic oscillator with Hamiltonian function%
\[
H=\frac{\omega_{x}}{2}(p_{x}^{2}+x^{2})+\frac{\omega_{y}}{2}(p_{y}^{2}%
+y^{2})\text{;}%
\]
we assume that the frequencies $\omega_{y}$, $\omega_{x}$ are incommensurate,
so that the only periodic orbits are librations along the $x$ and $y$ axes.
Let us focus on the orbit $\gamma_{x}$ along the $x$ axis; its prime period is
$T=2\pi/\omega_{x}$ and the corresponding monodromy matrix is%
\[
S_{1}=%
\begin{bmatrix}
1 & 0 & 0 & 0\\
0 & \cos\chi & 0 & \sin\chi\\
0 & 0 & 1 & 0\\
0 & -\sin\chi & 0 & \cos\chi
\end{bmatrix}
\text{ \ \ , \ }\chi=2\pi\frac{\omega_{y}}{\omega_{x}}\text{;}%
\]
it is the endpoint of the symplectic path $t\longmapsto S_{t}$, $0\leq t\leq
1$, consisting of the matrices
\[
S_{t}=%
\begin{bmatrix}
\cos2\pi t & 0 & \sin2\pi t & 0\\
0 & \cos\chi t & 0 & \sin\chi t\\
-\sin2\pi t & 0 & \cos2\pi t & 0\\
0 & -\sin\chi t & 0 & \cos\chi t
\end{bmatrix}
\text{.}%
\]
In Gutzwiller's formula (\ref{Gutz}) the sum is taken over periodic orbits,
including their repetitions; we are thus led to calculate the Conley--Zehnder
index of the path $t\longmapsto S_{t}$ with $0\leq t\leq r$ where the integer
$r$ indicates the number of repetitions of the orbit. Let us calculate the
Conley--Zehnder index $\nu(\tilde{S}_{r,\infty})$ of this path. We have
$S_{t}=\Sigma_{t}\oplus\tilde{S}_{t}$ where
\[
\Sigma_{t}=%
\begin{bmatrix}
\cos2\pi t & \sin2\pi t\\
-\sin2\pi t & \cos2\pi t
\end{bmatrix}
\text{ \ , \ }\tilde{S}_{t}=%
\begin{bmatrix}
\cos\chi t & \sin\chi t\\
-\sin\chi t & \cos\chi t
\end{bmatrix}
\text{;}%
\]
in view of the additivity property of the relative Maslov index we thus have
\[
\nu(S_{r,\infty})=\nu(\Sigma_{r,\infty})+\nu(\tilde{S}_{r,\infty})
\]
where the first term is just%
\[
\nu(\Sigma_{r,\infty})=-2r
\]
in view of the calculation we made in the one-dimensional case with a
different parametrization. Let us next calculate $\nu(\tilde{S}_{r,\infty})$.
We will use formula (\ref{muw}) relating the index $\nu$ to the Maslov index
via the index of concavity, so we begin by calculating the relative Maslov
index
\[
m_{\ell_{P}}(\tilde{S}_{r,\infty})=m(\tilde{S}_{r,\infty}\ell_{P,\infty}%
,\ell_{P_{,\infty}})\text{.}%
\]
Here is a direct argument; in more complicated cases the formulas we proved in
\cite{MSDG} are useful. When $t$ goes from $0$ to $r$ the line $\tilde{S}%
_{t}\ell_{P}$ describes a loop in $\operatorname*{Lag}(1)$ going from
$\ell_{P}$ to $\tilde{S}_{r}\ell_{P}$. We have $\tilde{S}_{t}\in U(1)$; its
image in $U(1,\mathbb{C})$ is $e^{-i\chi t}$ hence the Souriau mapping
identifies $\tilde{S}_{t}\ell_{P}$ with $e^{-2i\chi t}$. It follows, using
formula (\ref{souriau}), that
\begin{align*}
m_{\ell_{P}}(\tilde{S}_{r,\infty})  &  =\frac{1}{2\pi}\left(  -2r\chi
+i\operatorname{Log}(-e^{-2ir\chi})\right)  +\frac{1}{2}\\
&  =\frac{1}{2\pi}\left(  -2r\chi+i\operatorname{Log}(e^{i(-2r\chi+\pi
)})\right)  +\frac{1}{2}%
\end{align*}
The logarithm is calculated as follows: for $\theta\neq(2k+1)\pi$
($k\in\mathbb{Z}$)%
\[
\operatorname*{Log}e^{i\theta}=i\theta-2\pi i\left[  \frac{\theta+\pi}{2\pi
}\right]
\]
and hence%
\[
\operatorname{Log}(e^{i(-2r\chi+\pi)})=-i(2r\chi+\pi+2\pi\left[  \frac{r\chi
}{\pi}\right]  )\text{;}%
\]
it follows that the Maslov index is%
\begin{equation}
m_{\ell_{P}}(\tilde{S}_{r,\infty})=-\left[  \frac{r\chi}{\pi}\right]  \text{.}
\label{masell}%
\end{equation}
To obtain $\nu(\tilde{S}_{r,\infty})$ we note that by (\ref{muw})%
\[
\nu(\tilde{S}_{r,\infty})=m_{\ell_{P}}(\tilde{S}_{1,\infty}%
)-\operatorname*{Inert}W_{xx}^{\prime\prime}%
\]
where $\operatorname*{Inert}W_{xx}^{\prime\prime}$ is the concavity index
corresponding to the generating function of $\tilde{S}_{t}$; the latter is
\[
W(x,x^{\prime},t)=\frac{1}{2\sin\chi t}((x^{2}+x^{\prime2})\cos\chi
t-2xx^{\prime})
\]
hence $W_{xx}^{\prime\prime}=-\tan(\chi t/2)$. We thus have, taking
(\ref{masell}) into account,%
\[
\nu(\tilde{S}_{r,\infty})=-\left[  \frac{r\chi}{\pi}\right]
-\operatorname*{Inert}\left(  -\tan\frac{r\chi}{2}\right)  \text{;}%
\]
a straightforward induction on $r$ shows that this can be rewritten more
conveniently as%
\[
\nu(\tilde{S}_{r,\infty})=-1-2\left[  \frac{r\chi}{2\pi}\right]  \text{.}%
\]
Summarizing, we have%
\begin{align*}
\nu(S_{r,\infty})  &  =\nu(\Sigma_{r,\infty})+\nu(\tilde{S}_{r,\infty})\\
&  =-2r-1-2\left[  \frac{r\chi}{2\pi}\right]  )
\end{align*}
hence the index in Gutzwiller's formula corresponding to the $r$-th repetition
is%
\[
\mu_{x,r}=-\nu(S_{r,\infty})=1+2r+2\left[  \frac{r\chi}{2\pi}\right]
\]
that is, by definition of $\chi$,%
\[
\mu_{x,r}=1+2r+2\left[  r\frac{\omega_{y}}{\omega_{x}}\right]
\]
confirming the calculations in \cite{BB,CRL,BP,Sugita}.

\begin{remark}
The calculations above are valid when the frequencies are incommensurate. If,
say, $\omega_{x}=\omega_{y}$, the calculations are much simpler: in this case
the homotopy class of the loop $t\longmapsto S_{t}$, $0\leq t\leq1$, is
$\alpha^{-1}\oplus\alpha^{-1}$ and by the second formula (\ref{anu}),%
\[
\mu_{x,r}=-\nu(S_{r,\infty})=4r
\]
which is zero modulo $4$ (cf. \cite{BP}).
\end{remark}

\begin{acknowledgement}
One of the authors (M. de Gosson) has been supported by the grant 2005/51766-7
of the Funda\c{c}\~{a}o de Amparo \`{a} Pesquisa do Estado de S\~{a}o Paulo
(FAPESP); it is a pleasure for him to thank P Piccione for his kind
hospitality. \bigskip
\end{acknowledgement}


\begin{thebibliography}{99}                                                                                               %


\bibitem {BB}M Brack and R K Bhaduri 1997 \textit{Semiclassical Physics} (Addison--Wesley)

\bibitem {CLM}S E Cappell, R Lee and E Y Miller 1994 \textit{Comm. Pure and
Appl. Math.} \textbf{17 }121

\bibitem {CZ}C Conley and E Zehnder 1984 \textit{Comm. Pure and Appl. Math.
}\textbf{37} 207

\bibitem {CRL}S C Creagh, J M Robbins, and R\ G Littlejohn 1990 \textit{Phys.
Rev. A}, \textbf{4}2(4) 1907

\bibitem {AIF}M de Gosson\textit{\ }1990 \textit{Ann. Inst. Fourier}
\textbf{40}(3) 537

\bibitem {Cocycles}M de Gosson 1992 \textit{J. Geom. Phys. }\textbf{9} 255

\bibitem {JMPA}M de Gosson 1992 \textit{J. Math. Pures et Appl.} \textbf{71} 429

\bibitem {Wiley}M de Gosson 1997 \textit{Maslov Classes, Metaplectic
Representation and Lagrangian\ Quantization} (\textit{Research Notes in
Mathematics 95}\textbf{)} (Berlin: Wiley--VCH)

\bibitem {ICP}M de Gosson 2001 \textit{The Principles of Newtonian and Quantum
Mechanics: the Need for Planck's Constant }$h$\textit{; with a foreword by B.
Hiley} (London: Imperial College Press)

\bibitem {MSDG}M de Gosson and S de Gosson 2003 \textit{J. Phys A: Math. and
Gen.} \textbf{36} 615

\bibitem {mdglettmath}M de Gosson 2005 \textit{Letters in\ Mathematical
Physics }\textbf{72} 129

\bibitem {Gutzwiller}M C Gutzwiller 1990 \textit{Chaos in Classical and
Quantum Mechanics }(Interdisciplinary Applied Mathematics) (Berlin: Springer)

\bibitem {HWZ}H. Hofer, K Wysocki, and E Zehnder 1995 \textit{Geometric and
Functional Analysis} \textbf{2}(5) 270

\bibitem {Leray}J Leray 1981 \textit{Lagrangian Analysis and Quantum
Mechanics,\ a mathematical structure related to asymptotic expansions and the
Maslov index }(Cambridge, MA: MIT Press, )

\bibitem {LV}G Lion, and M Vergne 1980 \textit{The Weil representation, Maslov
index and Theta series} (\textit{Progress in mathematics} \textbf{6}) (Boston: Birkh\"{a}user)

\bibitem {WM}B Mehlig and M Wilkinson 2001 \textit{Ann. Phys}. \textbf{18}%
(10), 541

\bibitem {Morse}M Morse 1935 \textit{The Calculus of Variations in the Large}
(Providence, R I: American Mathemtical Society)

\bibitem {MG}P Muratore-Ginnaneschi 2003 \textit{Phys. Rep. }\textbf{383} 299

\bibitem {Piccione}R C Nostre Marques, P Piccione, and D Tausk 2001 in
\textit{Differential Geometry and its Applications} (Opava: Conference Proceedings)

\bibitem {BP}M Pletyukhov and M Brack 2003 \textit{J. Phys A: Math. and Gen.}
\textbf{36} 9449

\bibitem {Robbins}J M Robbin 1991 \textit{Nonlinearity} \textbf{4} 343

\bibitem {Souriau}J M Souriau 1975 In: Group Theoretical Methods in Physics,
Lecture Notes in Physics, Springer-Verlag, \textbf{50} 17

\bibitem {Sugita}A Sugita 2001 \textit{Ann. Phys.} \textbf{288 }277

\bibitem {Wall}C T C Wall 1969 \textit{Invent. Math.} \textbf{7} 269.
\end{thebibliography}
\end{document}